\RequirePackage{fix-cm}
\documentclass[smallextended]{svjour3}       % onecolumn (second format)
\smartqed  % flush right qed marks, e.g. at end of proof
\usepackage{graphicx}
\usepackage{amsmath}
 \usepackage{mathptmx}      % use Times fonts if available on your TeX system
\newcommand{\dirac}{\mathrm{d}}
\renewcommand{\d}{\mathrm{d}}
\newcommand{\glv}{\gamma_\mathrm{LV}}
\newcommand{\gsl}{\gamma_\mathrm{SL}}
\newcommand{\gsv}{\gamma_\mathrm{SV}}
\begin{document}

\title{Elastocapillary coiling of an elastic rod inside a drop
}

%\titlerunning{-}        % if too long for running head

\author{Herv\'e Elettro \and Paul Grandgeorge  \and S\'ebastien Neukirch}

\institute{
Sorbonne Universit\'es, UPMC Univ Paris 06, CNRS, UMR 7190, Institut Jean Le Rond d'Alembert, F-75005 Paris, France
\at
\email{sebastien.neukirch@upmc.fr} 
}

\date{Received: date / Accepted: date}

\maketitle
\begin{abstract}
Capillary forces acting at the surface of a liquid drop can be strong enough to deform small objects and recent studies have provided several examples of elastic instabilities induced by surface tension.
We present such an example where a liquid drop sits on a straight fiber, and we show that the liquid attracts the fiber which thereby coils inside the drop.
We derive the equilibrium equations for the system, compute bifurcation curves, and show the packed fiber may adopt several possible configurations inside the drop.
We use the energy of the system to discriminate  between the different configurations and find a intermittent regime between two-dimensional  and three-dimensional solutions as more and more fiber is driven inside the drop.
\end{abstract}

\section{Introduction} \label{intro}
%=======================
%
%
%
%
The packaging of elastic filaments in cavities \cite{Stoop2011} is a model system for a large variety of physical phenomena, for example the ejection of DNA from viral capsids \cite{Leforestier2009Structure-of-toroidal,lamarque+harvey:2004,Katzav2006}, carbone nanotubes compaction \cite{Chen2013General-Methodology}, or the windlass mechanism in spider capture threads \cite{Vollrath1989}.
In the case of a mechanical wire spooled in a sphere, or DNA in a capsid, the presence of a motor is necessary for the packing process, the energy to bend the filament being provided by this external actuator. In the case the cavity is a liquid drop (or a bubble in a liquid medium) surface tension may provide the actuation energy: if the affinity of the filament for the liquid is stronger than that of the filament for the surrounding gas, then the bending energy required for packing could be provided by the difference of surface energies.
As always, surface energy prevails at small scale and carbon nanotubes adopting ring shapes in cavitation bubbles have been experimentally observed \cite{Martel1999Rings-of-single-walled} and theoretically analyzed \cite{cohen+mahadevan:2003}. For larger drop-on-fiber systems \cite{Lorenceau2004Capturing-drops,Duprat2012Wetting-of-flexible}, typically of millimeter or centimeter sizes, the competition between capillary and elastic forces is not automatically won by the former: a threshold length emerges \cite{cohen+mahadevan:2003} and separates systems in which packaging is possible from those in which the fiber remains straight.
This threshold length, called elastocapillary length \cite{Roman2010}, plays a central role in problems in which surface tension bends or buckles slender rods \cite{Fargette2014} or thin elastic sheets \cite{Py2007}.

Computations of configurations of a filament  packaged in a spherical cavity have been performed using Finite Elements \cite{Vetter2013}, molecular mechanics \cite{arsuaga+al:2002}, or statistical physics \cite{Adda-Bedia2010Statistical-distributions} approaches.
Here we present a model for the buckling and coiling of an elastic rod in a rigid spherical cavity. In section \ref{section:model} we derive the equilibrium equations for the elastic rod using an energy approach \cite{Steigmann-1993,Bourgat-1988}. In Section \ref{section:bvp} we non-dimensionalize the equations and set the boundary-value problem which we numerically solve. In Section \ref{section:bif-diag} we plot bifurcation curves and compare the energy of the different coiling solutions. We finally present experimental  configurations of coiled systems involving elastomeric beams and oil droplets.

\section{Surface tension and interface energy}\label{section:capi}
%=====================================
%
%
%
%
%
%
%
%
%
\begin{figure}
\center
\includegraphics[width=0.5\textwidth]{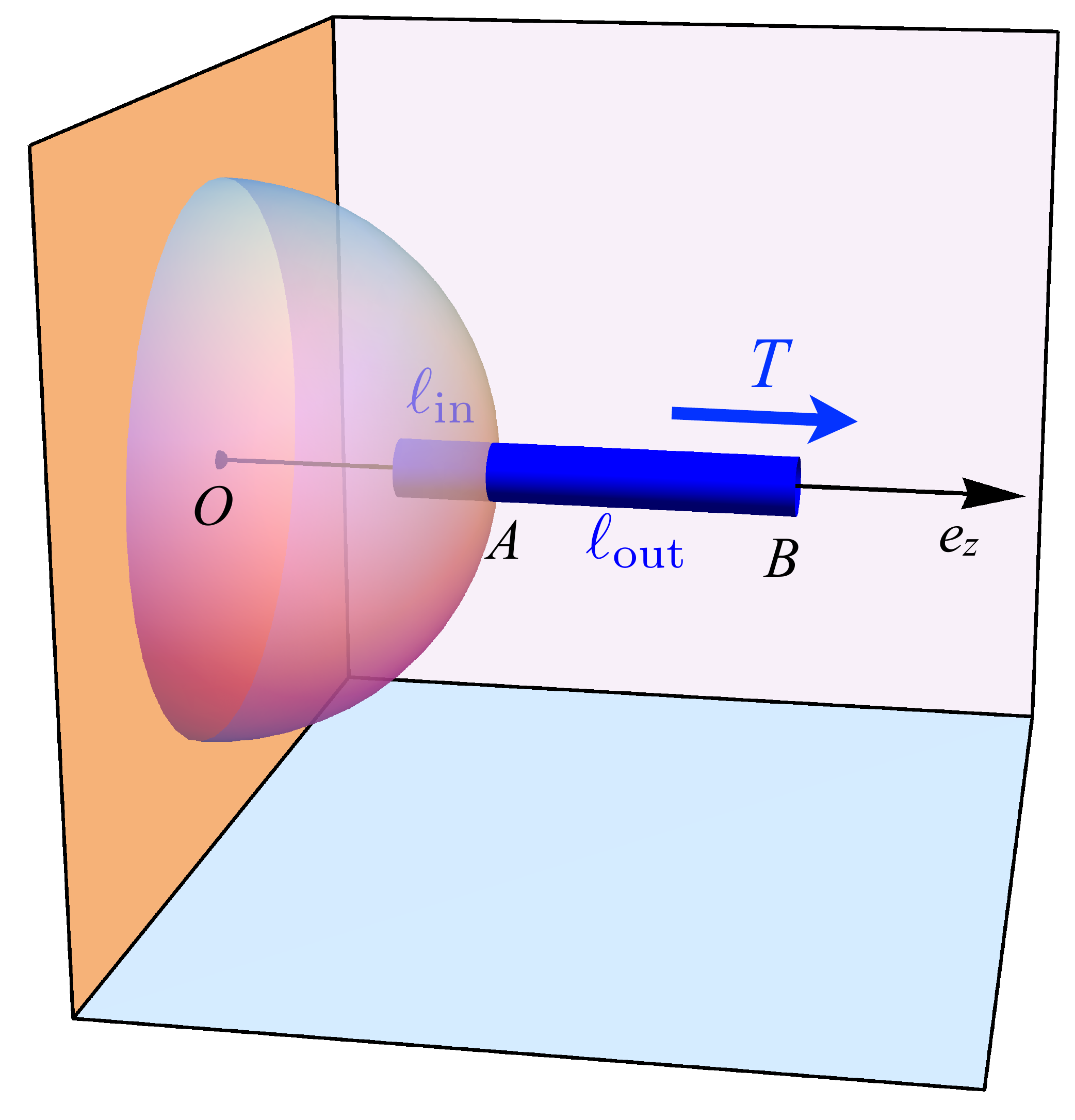}
\caption{A cylindrical rigid rod partially immersed in a liquid drop. As affinity of the rod for the liquid is stronger than that with air, a tension $T$ is applied to prevent complete immersion. A capillary force, applied by the liquid at point $A$, is balancing this tension $T$.}
\label{fig:capi}
\end{figure}
In this section we recall the link between surface tension and surface energy. Any interface between two mediums costs energy as molecules lying at the interface are in a less favorable state than molecules deeply buried in the medium. 
Theses molecules at the interface are then slightly scattered and therefore in a state of tension: surface energy yields surface tension \cite{degennes+al:2003}. We illustrate this link in Figure~\ref{fig:capi} which shows a drop sitting on a wall in the absence of gravity. A cylindrical rigid rod is then partially immersed in the liquid drop. The rod's material has a stronger affinity with the liquid than with air and consequently the rod is attracted toward the liquid. One has then to pull the rod at its right end to equilibrate the system.
We compute this pulling tension $T$ by writing the potential energy of the system. The solid-liquid interface has energy $\gsl$ per unit area, yielding a total surface energy $\ell_\text{in} \, P \, \gsl + \Sigma \, \gsl$ where $P$ is the perimeter and $\Sigma$ the area of the cylinder cross-section, and $\ell_\text{in}$ the immersed length. Similarly the total surface energy of the emerged part (of length $\ell_\text{out}$) is $\ell_\text{out} \, P \, \gsv + \Sigma \, \gsv$. The liquid-vapor surface energy is $2 \pi \, R^2 \glv$ where $R$ is the (constant) radius of the liquid drop. Summing up all these energies, adding the work $-T \, z(B)$ of the external tension $T$, and dropping out constant terms, we write the total potential energy of the system as
\begin{equation}
E=\ell_\text{in} \, P \, \gsl + \ell_\text{out} \, P \, \gsv - T \, z(B) \, .
\end{equation}
Replacing $\ell_\text{in}= L - \ell_\text{out}$ andv $z(B) = R + \ell_\text{out}$ we finally arrive at
\begin{equation}
E(\ell_\text{out}) = \ell_\text{out} \left( P [\gsv-\gsl] -T \right) + \mbox{const.}
\end{equation}
Equilibrium is then achieved for $\d E / \d \ell_\text{out}=0$ that is
\begin{equation}
T = P \, (\gsv-\gsl) >0 \, .
\end{equation}
In the language of forces we say that the external tension $T$ is balanced by a capillary force $F_\gamma = P (\gsv-\gsl)$, applied by the liquid on the rod at point $A$ and oriented toward the center of the drop.

\section{Variational approach}\label{section:model}
%=====================================
%
%
%
%
%
%
%
%
%

We consider a rod that is naturally straight and has a linear elastic response to bending and twisting. In addition, 
we work under the assumption that the rod is inextensible and unshearable.
The position of the center line of the rod is $\mathbf{R}(S)$
where  $S$ is the arc length along the rod. The arc length runs from $0$ to $L$, $L$ being the total contour length of the rod. 
To follow the deformation of the rod material around the center line, we use a set of Cosserat orthonormal directors $\{\mathbf{d_1}(S), \mathbf{d_2}(S), \mathbf{d_3}(S)\}$ where $\mathbf{d_3}(S)$ is the tangent to the rod center line and $\{\mathbf{d_1}(S),\mathbf{d_2}(S)\}$ register the rotation of the rod cross section about the tangent.
%\begin{equation}
%\mathbf{R}'(S) = \mathbf{d_3}(S)
%\end{equation}
Orthonormality of the Cosserat frame implies the existence of a Darboux vector $\mathbf{U}(S)$ such that 
\begin{equation}
\label{equa:darboux}
\mathbf{d_i}'(S) = \mathbf{U}(S) \times \mathbf{d_i}(S) \; , \quad i=1,2,3.
\end{equation}
The components $U_i =\mathbf{U} \cdot \mathbf{d_i}$ are used to write the strain energy of the rod
\begin{equation}
E_\mathrm{strain} = \int_0^L 
%\frac{1}{2} 
(1/2)
\left[ K_1 \, U_1^2(S) + K_2 \, U_2^2(S) + K_3 \, U_3^2(S)  \right] \d S \, , 
\end{equation}
where  $K_1$ and $K_2$ are the bending rigidities, $K_3$ is the twist rigidity, $U_1$ and $U_2$ are the curvature strains, and $U_3$ is the twist strain\cite{Antman2004,Audoly-Pomeau2010}.

\begin{figure}
\center
\includegraphics[width=0.95\textwidth]{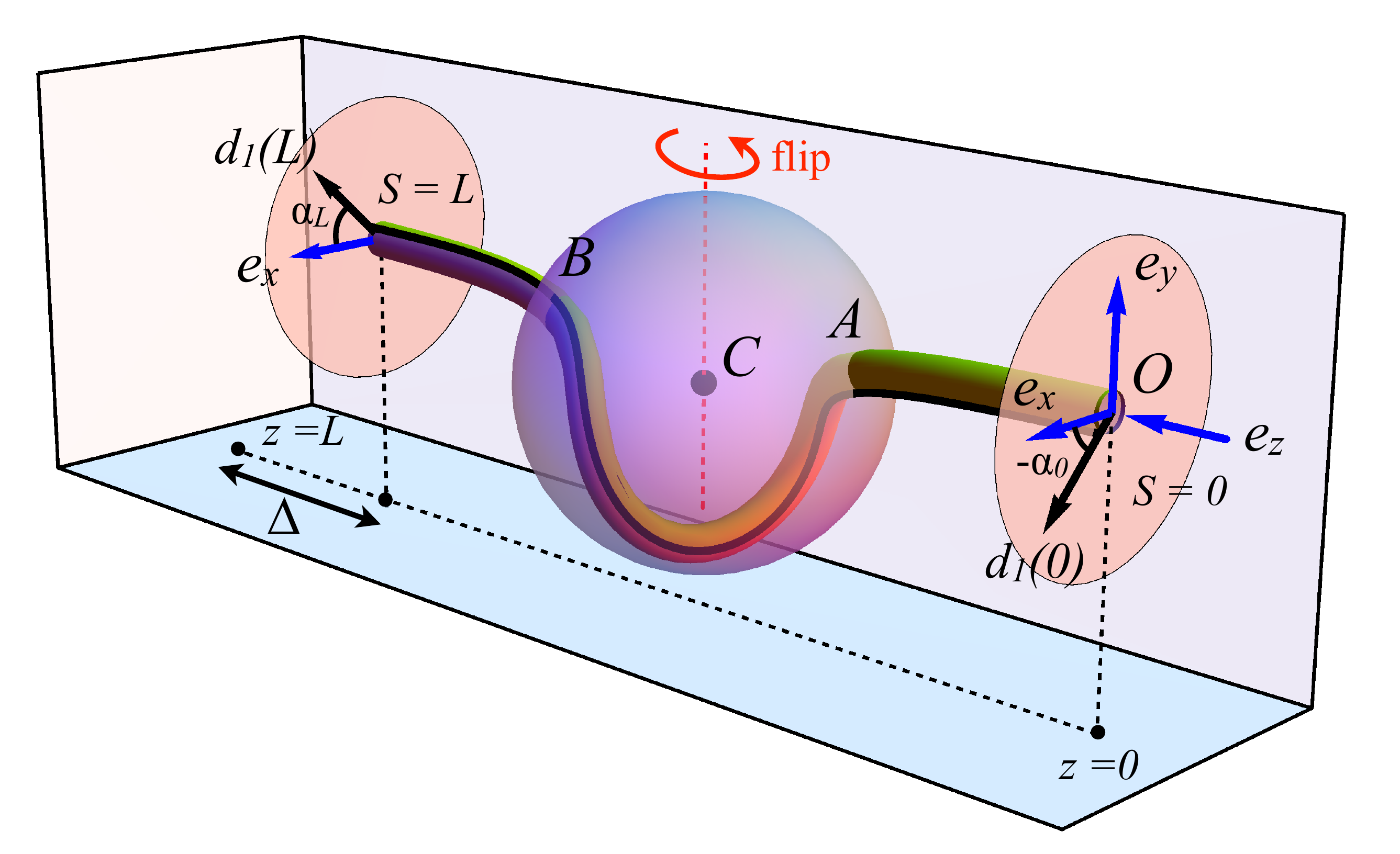}
\caption{Equilibrium configuration of an elastic rod buckled by a liquid spherical drop. The rod is clamped at both extremities, and an end-shortening $\Delta$ is imposed. The configuration is symmetric about a $\pi$-rotation about the axis joining the center $C$ of the sphere and the rod middle point $S=L/2$. Meniscus forces, applied on the rod at meniscus points $A$ and $B$, together with soft-wall barrier forces are represented. Their sum is equal to the weight of the drop, see Equation (\ref{equa:cequivient-de-la-goutte}). }
\label{fig:configuration}
\end{figure}
A liquid drop of mass $M$ and radius $R$ is attached to the rod. The rod enters the drop at meniscus point $A$ and exits the drop at meniscus point $B$. Writing $S_A$ the arc length of point $A$ and $S_B$ the arc length of point $B$, 
we divide the rod in three regions: (I) where $S\in [0;S_A)$, (II) where $S\in (S_A;S_B)$, and (III) where $S\in (S_B;L]$, with region (II) lying inside the liquid drop.
In regions (I) and (III) the solid-vapor interface has an energy $\gsv$ per unit area. In region (II) the solid-liquid interface has energy $\gsl$ per unit area. The total surface energy of the rod then scales linearly with the contour length:
\begin{equation}
E_\mathrm{surface} = P \left[  \gsv S_A +\gsl (S_B-S_A) +\gsv (L-S_B) \right] \, , 
\end{equation}
where $P$ is the perimeter of the cross-section of the rod.
If the drop was free-standing it would adopt a spherical shape in order to minimize its surface energy.
Nevertheless due to $(i)$ its own weight, $(ii)$ contact pressure coming from the coiled rod, and $(iii)$ meniscus contact angles, the drop is non-spherical. But as we deal with $(i)$ drops which are smaller than the capillary length,  $(ii)$ rods which are flexible compared to surface tension, and $(iii)$ drops with radii much larger than the diameter of the cross-section of the rod, to a first approximation we consider the drop to be spherical.
The surface energy $4 \pi R^2 \glv$ corresponding to the liquid-vapor interface ({\em i.e.} the drop surface) is then constant and therefore discarded.
As in \cite{Elettro2015} we use a soft-wall potential to prevent the rod from exiting the sphere elsewhere than at the meniscus points $A$ and $B$:
\begin{equation}
V_\mathrm{wall}\left(\mathbf{R}(S),\mathbf{R_C}\right) = \frac{V_0}{1 + \rho -(1/R) \sqrt{(\mathbf{R}(S)-\mathbf{R_C})^2}} \, ,
\end{equation}
where $\mathbf{R_C}=(X_C,Y_C,Z_C)^T$ is the position of the center of the spherical drop, and $V_0$ is an energy per unit length with $V_0 \to 0$ corresponding to the hard-wall limit.
The small dimensionless parameter $\rho$ is introduced to avoid the potential to diverge at the meniscus points $A$ and $B$, where the rod enters and exits the sphere.
As we are working with sub-millimetric systems we neglect the weight or the rod, but take into account the weight of the drop by adding the term $M g Y_C$ into the energy. 
The total potential energy of the system is then, up to constant terms,
\begin{equation}
E_\mathrm{tot}=E_\mathrm{strain} + \int_{S_A}^{S_B} V_\mathrm{wall} \, \d S -F_\gamma (S_B-S_A) + M g Y_C \, ,
\end{equation}
where we have introduced the surface tension force $F_\gamma = P (\gsv - \gsl) > 0$.
We consider boundary conditions of clamping, where the rod position $\mathbf{R}(S)$ and rotation are restrained at both sides:
\begin{subequations}
\label{sys:BC}
\begin{align}
\mathbf{R}(0)=\mathbf{0} \; &,  \: \:  \mathbf{R}(L)=(0,0,L-\Delta)^T  \, , \\
\mathbf{d_1}(0)=\cos \alpha_0 \,  \mathbf{e_x} + \sin \alpha_0 \, \mathbf{e_y}  \; &, \: \: 
\mathbf{d_2}(0)=-\sin \alpha_0 \,  \mathbf{e_x} + \cos \alpha_0 \, \mathbf{e_y}  \; , \: \: 
\mathbf{d_3}(0)=\mathbf{e_z} \, , \\
\mathbf{d_1}(L)=\cos \alpha_L \,  \mathbf{e_x} + \sin \alpha_L \, \mathbf{e_y} \; &, \: \: 
\mathbf{d_2}(L)=-\sin \alpha_L \,  \mathbf{e_x} + \cos \alpha_L \, \mathbf{e_y}   \; , \: \: 
\mathbf{d_3}(L)=\mathbf{e_z} \, ,
\end{align}
\end{subequations} 
where $\Delta$ is the prescribed end-shortening, $\alpha_0$ and $\alpha_L$ are fixed angles, and $\{\mathbf{e_x}, \mathbf{e_y}, \mathbf{e_z} \}$ is our reference frame.
We minimize $E_\mathrm{tot}$ under the following constraints
\begin{subequations}
\label{sys:constraints}
\begin{align}
\mathbf{R}'(S)=\mathbf{d_3}(S)  \quad &\text{(inextensibility)} \label{equa:inext-constr} \, , \\
\mathbf{d_i}(S) \cdot \mathbf{d_j}(S) = \delta_{ij} \quad &\text{(orthonormality)} \, , \\
U_1 = \mathbf{d_2}' \cdot \mathbf{d_3}   \; , \quad U_2 = \mathbf{d_3}' \cdot \mathbf{d_1}
  \; , \quad U_3 = \mathbf{d_1}' \cdot \mathbf{d_2} \quad &\text{(Darboux relations)} \, , \\
\left( \mathbf{R}(S_A) - \mathbf{R_C} \right)^2 = R^2 = \left( \mathbf{R}(S_B) - \mathbf{R_C} \right)^2  \quad &\text{(position of meniscus)} \, , \label{equa:AetBsphere}
\end{align}
\end{subequations} 
where $' \equiv \d/ \d s$. We cope with constraints by introducing Lagrange multipliers and considering the Lagrangian
\begin{align}
{\cal L}&(U_1,U_2,U_3,\mathbf{R},\mathbf{d_1},\mathbf{d_2},\mathbf{d_3},S_A,S_B,\mathbf{R_C})= \nonumber \\ 
&\int_0^L \Big[ (1/2)  \left( K_1 \, U_1^2 + K_2 \, U_2^2 + K_3 \, U_3^2  \right)
-  M_1 \left( U_1 - \mathbf{d_2}' \cdot \mathbf{d_3} \right)  
- M_2 \left( U_2 - \mathbf{d_3}' \cdot \mathbf{d_1} \right)   \nonumber \\
& -M_3 \left(U_3 - \mathbf{d_1}' \cdot \mathbf{d_2} \right) -\nu_{12} \left( \mathbf{d_1} \cdot \mathbf{d_2} \right)
-\nu_{13} \left( \mathbf{d_1} \cdot \mathbf{d_3} \right)
-\nu_{23} \left( \mathbf{d_2} \cdot \mathbf{d_3} \right) \nonumber \\
& -(1/2) \epsilon_1 \left( \mathbf{d_1} \cdot \mathbf{d_1} - 1\right)
-(1/2) \epsilon_2 \left( \mathbf{d_2} \cdot \mathbf{d_2} - 1\right)
-(1/2) \epsilon_3 \left( \mathbf{d_3} \cdot \mathbf{d_3} - 1\right) \Big] \, \d S \nonumber \\
& + (F_A / 2R) \left[  \left( \mathbf{R}(S_A) - \mathbf{R_C} \right)^2 -R^2\right]
+ (F_B / 2R) \left[  \left( \mathbf{R}(S_B) - \mathbf{R_C} \right)^2 -R^2\right]  \nonumber \\
& -F_\gamma \, (S_B-S_A) - \mathbf{W} \cdot  \mathbf{R_C} 
+ \int_{S_A}^{S_B} V_\mathrm{wall} \, \d S  \nonumber \\
&+ \int_0^{S_A} \mathbf{N}^\mathrm{I} \cdot \left( \mathbf{R}' - \mathbf{d_3}\right) \d S
+ \int_{S_A}^{S_B} \mathbf{N}^\mathrm{II} \cdot \left( \mathbf{R}' - \mathbf{d_3}\right) \d S
+ \int_{S_B}^{L} \mathbf{N}^\mathrm{III} \cdot \left( \mathbf{R}' - \mathbf{d_3}\right) \d S \, , 
\end{align}
where $\mathbf{W}=(0,-Mg,0)^T$ is the weight of the drop, and where $M_i(S)$, $\nu_{ij}(S)$, $\epsilon_i(S)$, $\mathbf{N}^\mathrm{I,II,III}(S)$, and $F_{A,B}$ are Lagrange multipliers.
As we will see in the following, the multipliers $M_i(S)$ correspond to the components of the internal moment $\mathbf{M}$, $\mathbf{M}=M_1 \mathbf{d_1}+M_2 \mathbf{d_2}+M_3 \mathbf{d_3}$. It can be shown that, as in the planar case \cite{Elettro2015}, $\mathbf{M}(S)$ does not experience any discontinuity at the meniscus points $S=S_A$ and $S=S_B$. 
In the same manner, the multipliers $\mathbf{N}^\mathrm{I,II,III}(S)$ will be interpreted as the internal force in each region of the system.
Since the internal force in the rod does experience discontinuities at the meniscus points, we have therefore introduced three different multipliers $\mathbf{N}^\mathrm{I,II,III}(S)$.

\paragraph{First variation} ~ \\
We note $\mathbf{w}=(U_1,U_2,U_3,\mathbf{R},\mathbf{d_1},\mathbf{d_2},\mathbf{d_3},S_A,S_B,\mathbf{R_C})$ and we consider the conditions for a state $\mathbf{w}_e$ to minimize the energy $E_\mathrm{tot}$. Calculus of variations shows that a necessary condition is 
\begin{equation}
{\cal L}'(\mathbf{w}_e) \cdot  \overline{\mathbf{w}} =
\left. \frac{\d}{\d \epsilon} {\cal L}(\mathbf{w}_e + \epsilon  \overline{\mathbf{w}}) \right|_{\epsilon=0}= 0 \;  \quad \forall \,\overline{\mathbf{w}} \, , 
\label{equa:def-1st-varia}
\end{equation}
where $\overline{\mathbf{w}}$ is a variation of the variable ${\mathbf{w}}$.
% PAS BESOIN , and $|\epsilon| \ll 1$.
%
Boundary conditions (\ref{sys:BC}) imply that
\begin{align}
\overline{\mathbf{R}}(0)=\mathbf{0}       \; , & \quad  \overline{\mathbf{R}}(L) =\mathbf{0}  \label{equa:BC_bar_R} \, , \\
\overline{\mathbf{d_i}}(0,L)=\mathbf{0}  \; , & \quad \overline{\mathbf{d_i}}(L)=\mathbf{0} \; \quad \forall i  \, , \label{equa:BC_bar_D}
\end{align} 
Noting that $\left.  \frac{\d}{\d \epsilon} \: \int_0^{a+\epsilon \overline a} f(x) \, \d x \: \: \right|_{\epsilon=0}= \overline a \, f(a) $ we calculate the first variation (\ref{equa:def-1st-varia})
%
%
%Noting that $\int_0^{a+\epsilon \overline a} f(x) \, \d x = \int_0^a f(x) \, \d x + \epsilon \overline a \, f(a) + O(\epsilon^2)$ we calculate the first variation (\ref{equa:def-1st-varia})
%
\begin{align}
{\cal L}'(\mathbf{w}_e) & \cdot  \overline{\mathbf{w}} = \nonumber \\
&\int_0^L \Big[ K_1 \, U_1 \, \overline U_1 + K_2 \, U_2 \, \overline U_2 + K_3 \, U_3 \,\overline U_3  \nonumber \\
&-  M_1 \left( \overline{U_1} -  \overline{\mathbf{d_2}}' \cdot \mathbf{d_3} - \mathbf{d_2}' \cdot  \overline{\mathbf{d_3}}  \right)  
- M_2 \left( \overline{U_2} - \overline{\mathbf{d_3}}' \cdot \mathbf{d_1}- \mathbf{d_3}' \cdot \overline{\mathbf{d_1}} \right)  \nonumber \\
& -M_3 \left(\overline{U_3} - \overline{\mathbf{d_1}}' \cdot \mathbf{d_2}- \mathbf{d_1}' \cdot \overline{\mathbf{d_2}} \right) 
-\nu_{12} \left( \overline{\mathbf{d_1}} \cdot \mathbf{d_2} + \mathbf{d_1} \cdot \overline{\mathbf{d_2}}  \right) \nonumber \\
&-\nu_{13} \left(  \overline{\mathbf{d_1}} \cdot \mathbf{d_3} +  \mathbf{d_1} \cdot  \overline{\mathbf{d_3}} \right)
-\nu_{23} \left( \overline{\mathbf{d_2}} \cdot \mathbf{d_3} + \mathbf{d_2} \cdot \overline{\mathbf{d_3}} \right) \nonumber \\
& - \epsilon_1  \overline{\mathbf{d_1}} \cdot \mathbf{d_1}
- \epsilon_2  \overline{\mathbf{d_2}} \cdot \mathbf{d_2} 
- \epsilon_3  \overline{\mathbf{d_3}} \cdot \mathbf{d_3}  \Big] \, \d S \nonumber \\
& + (F_A / R) \left[  \left( \mathbf{R}(S_A) - \mathbf{R_C} \right) \cdot \left( \overline{\mathbf{R}}(S_A)+ \overline{S_A} \; \mathbf{R}'(S_A) - \overline{\mathbf{R_C}} \right) \right] \nonumber \\
&+ (F_B / R) \left[  \left( \mathbf{R}(S_B) - \mathbf{R_C} \right) \cdot \left( \overline{\mathbf{R}}(S_B)+ \overline{S_B} \; \mathbf{R}'(S_B) -  \overline{\mathbf{R_C}} \right)   \right]  \nonumber \\
& -F_\gamma \, (\overline{S_B}-\overline{S_A}) - \mathbf{W} \cdot  \overline{\mathbf{R_C}}  \nonumber \\
&+ \int_{S_A}^{S_B} \left( \frac{\partial V_\mathrm{wall}}{\partial \mathbf{R}}  \cdot \overline{\mathbf{R}} + \frac{\partial V_\mathrm{wall}}{\partial \mathbf{R_C}}  \cdot \overline{\mathbf{R_C}} \right) \, \d S  
+ \overline{S_B} \, V_\mathrm{wall}(S_B) -  \overline{S_A} \, V_\mathrm{wall}(S_A)  \nonumber \\
&+ \int_0^{S_A} \mathbf{N}^\mathrm{I} \cdot \left( \overline{\mathbf{R}}' - \overline{\mathbf{d_3}}\right) \d S
+ \int_{S_A}^{S_B} \mathbf{N}^\mathrm{II} \cdot \left( \overline{\mathbf{R}}' - \overline{\mathbf{d_3}}\right) \d S
+ \int_{S_B}^{L} \mathbf{N}^\mathrm{III} \cdot \left( \overline{\mathbf{R}}' - \overline{\mathbf{d_3}}\right) \d S \, . 
\label{equa:1st-varia}
\end{align}
Note that we have used (\ref{equa:inext-constr}) to eliminate several terms.
We first require (\ref{equa:1st-varia}) to vanish for all $\overline{U_i}$ and obtain
\begin{equation}
M_i = K_i \, U_i  \, . \label{equa:const-rel}
\end{equation}
Equation (\ref{equa:const-rel}) appears as the bending/twisting constitutive relation of the elastic rod provided the Lagrange multipliers $M_i$ are seen as the components of the internal moment.
We next perform integrations by part for the terms involving $\overline{\mathbf{d_i}}'$. Boundary conditions (\ref{equa:BC_bar_D}) implies that the boundary terms vanish. Requiring the result to vanish for all $\overline{\mathbf{d_i}}$ yields:
\begin{subequations}
\label{sys:equa-di-bar}
\begin{align}
M_2 \, \mathbf{d_3}' - M_3 '  \, \mathbf{d_2} - M_3 \, \mathbf{d_2}' - \nu_{12} \, \mathbf{d_2} - \nu_{13} \, \mathbf{d_3} - \epsilon_1 \, \mathbf{d_1} = & \, \mathbf{0} \, ,  \\
M_3 \, \mathbf{d_1}' - M_1 '  \, \mathbf{d_3} - M_1 \, \mathbf{d_3}' - \nu_{23} \, \mathbf{d_3} - \nu_{12} \, \mathbf{d_1} - \epsilon_2 \, \mathbf{d_2} = & \, \mathbf{0} \, ,  \\
- \mathbf{N}^\mathrm{I,II,III} +  M_1 \, \mathbf{d_2}' - M_2 '  \, \mathbf{d_1} - M_2 \, \mathbf{d_1}' - \nu_{13} \, \mathbf{d_1} - \nu_{23} \, \mathbf{d_2} - \epsilon_3 \, \mathbf{d_3} = & \, \mathbf{0}  \, . 
\end{align}
\end{subequations}
We take the scalar product of each of these three equations with ${\mathbf{d_1}}$, $\mathbf{d_2}$, and $\mathbf{d_3}$ and combine them to eliminate the $\nu_{ij}$ and the $\epsilon_i$ to finally obtain
\begin{subequations}
\label{sys:moment-balance}
\begin{align}
M_1' = & M_2 \, U_3 - M_3 \, U_2 + N^\mathrm{I,II,III}_2 \, , \\
M_2' = & M_3 \, U_1 - M_1 \, U_3 - N^\mathrm{I,II,III}_1 \, , \\
M_3' = & M_1 \, U_2 - M_2 \, U_1  \, ,  \label{equa_M3}
\end{align}
\end{subequations}
which is the component version of the moment balance equation for the elastic rod, $\mathbf{M}' = \mathbf{N} \times \mathbf{R}'$.
The Lagrange multiplier $\mathbf{N}$ is then seen as the internal force in the rod.
We now perform integrations by parts for the terms involving $\overline{\mathbf{R}}'$ and find that
\begin{align}
& \, (F_A/R)  \left( \mathbf{R}(S_A) - \mathbf{R_C} \right) \cdot  \overline{\mathbf{R}}(S_A) +
(F_B/R)  \left( \mathbf{R}(S_B) - \mathbf{R_C} \right) \cdot  \overline{\mathbf{R}}(S_B)  \nonumber \\
&+ \int_{S_A}^{S_B}  \frac{\partial V_\mathrm{wall}}{\partial \mathbf{R}}  \cdot \overline{\mathbf{R}} 
+\left[   \mathbf{N}^\mathrm{I} \cdot  \overline{\mathbf{R}} \,  \right]_0^{S_A}
+\left[   \mathbf{N}^\mathrm{II} \cdot  \overline{\mathbf{R}}  \, \right]_{S_A}^{S_B}
+\left[   \mathbf{N}^\mathrm{III} \cdot  \overline{\mathbf{R}} \,  \right]_{S_B}^L \nonumber \\
&- \int_0^{S_A} \mathbf{N}^\mathrm{I}(S)' \cdot  \overline{\mathbf{R}}  \, \d S
- \int_{S_A}^{S_B} \mathbf{N}^\mathrm{II}(S)' \cdot  \overline{\mathbf{R}} \, \d S
- \int_{S_B}^{L} \mathbf{N}^\mathrm{III}(S)' \cdot  \overline{\mathbf{R}} \, \d S \; = 0 \quad \forall \, \overline{\mathbf{R}} \, . 
\end{align}
This condition implies the balance equations for the rod:
\begin{subequations}
\begin{align}
\mathbf{N}^\mathrm{I}(S)' &= \mathbf{0} \quad \forall S \in [0;S_A) \, , \\
\mathbf{N}^\mathrm{II}(S)' -  \frac{\partial V_\mathrm{wall}}{\partial \mathbf{R}} &= 
\mathbf{0} \quad \forall S \in (S_A;S_B) \, , \label{equa:force-balance-inside-drop}\\
\mathbf{N}^\mathrm{III}(S)' &= \mathbf{0} \quad \forall S \in (S_B;L] \, , 
\end{align}
\end{subequations}
where $\mathbf{N}^\mathrm{I,II,III}$ are  the internal forces in each region of the system.
The force per unit length $- \partial V_\mathrm{wall}/\partial \mathbf{R}$ corresponds to the soft-wall repulsion, from the drop interface, on the rod.
Boundary conditions (\ref{equa:BC_bar_R}) make the boundary terms at $S=0$ and $S=L$ vanish, but arbitrariness of the 
variations $\overline{\mathbf{R}}(S_A)$ and $\overline{\mathbf{R}}(S_B)$ require that
\begin{subequations}
\label{sys:meniscus-force}
\begin{align}
\mathbf{N}^\mathrm{II}(S_A) - \mathbf{N}^\mathrm{I}(S_A) + F_A \, \left[ \mathbf{R_C} - \mathbf{R}(S_A) \right]/R &= \mathbf{0}  \, , \\
\mathbf{N}^\mathrm{III}(S_B) - \mathbf{N}^\mathrm{II}(S_B) + F_B \, \left[ \mathbf{R_C} - \mathbf{R}(S_B ) \right]/R &= \mathbf{0}  \, . 
\end{align}
\end{subequations}
These equations correspond to forces balance at the meniscus points. We see that at meniscus points $A$ and $B$, the total force from the drop on the rod is oriented toward {\em the center of the spherical drop} and has intensity $F_{A}$ and $F_{B}$ respectively.
Requiring  (\ref{equa:1st-varia}) to vanish for all $\overline{\mathbf{R_C}}$ yields
\begin{equation}
\label{equa:cequivient-de-la-goutte}
F_A \, \left[ \mathbf{R_C} - \mathbf{R}(S_A) \right]/R + F_B \, \left[ \mathbf{R_C} - \mathbf{R}(S_B ) \right]/R = \mathbf{W}
+ \int_{S_A}^{S_B} \partial V_\mathrm{wall} / \partial \mathbf{R} \, \d S \, , 
\end{equation}
where we have used the identity $\partial V_\mathrm{wall}/\partial \mathbf{R_C}=-\partial V_\mathrm{wall}/\partial \mathbf{R}$.
Equation (\ref{equa:cequivient-de-la-goutte}) tells us that the total meniscus force from the drop on the rod (the right hand side of (\ref{equa:cequivient-de-la-goutte})) is equal to the weight $\mathbf{W}$ of the drop plus the opposite of the integrated soft-wall repulsion. We therefore see that the soft-wall repulsion applied inside the drop is balanced by the meniscus force. Accordingly, using (\ref{equa:force-balance-inside-drop}), (\ref{sys:meniscus-force}), and (\ref{equa:cequivient-de-la-goutte}) we find
\begin{equation}
\mathbf{N}^\mathrm{III}(S_B) - \mathbf{N}^\mathrm{I}(S_A) + \mathbf{W}  = \mathbf{0} \, . 
\label{equa:saut_global}
\end{equation}
That is, the total force from the liquid on regions I and III of the rod is simply the weight of the drop.
Finally the conditions for  (\ref{equa:1st-varia}) to vanish for all $\overline{S_A}$ and $\overline{S_B}$ are
\begin{subequations}
\label{sys:resolveFAB}
\begin{align}
(F_A/R) \, \left[ \mathbf{R_C} - \mathbf{R}(S_A) \right] \cdot \mathbf{R}'(S_A) &= F_\gamma - V_\mathrm{wall}(S_A) \, ,  \\
(F_B/R) \, \left[ \mathbf{R_C} - \mathbf{R}(S_B) \right] \cdot \mathbf{R}'(S_B) &= -F_\gamma + V_\mathrm{wall}(S_B) \, .\label{equa:fb}
\end{align}
\end{subequations}
These conditions can be seen as a way to compute the intensity $F_A$ and $F_B$ of the meniscus force. In particular we see that 
$F_A$ ($F_B$) depends on the relative orientation of the rod's tangent and the radial vector at the meniscus points $A$ ($B$), as also shown in \cite{Neukirch2007}.
Introducing the tension in rod $T(S) := \mathbf{N} \cdot \mathbf{d_3}$ and using (\ref{sys:meniscus-force}), we find that
\begin{subequations}
\label{sys:tension-jump}
\begin{align}
T^\mathrm{I}(S_A)  - T^\mathrm{II}(S_A)  &= F_\gamma - V_\mathrm{wall}(S_A) \, ,  \\
T^\mathrm{III}(S_B)  - T^\mathrm{II}(S_B)  &= F_\gamma - V_\mathrm{wall}(S_B) \, , 
\end{align}
\end{subequations}
where we see that, in the limit $V_0 \to 0$, the jump in the rod's tension is equal to the surface tension force $F_\gamma$.

\section{Boundary value problem} \label{section:bvp}
%==============
%
%
%
%
%
%
%
%
%
%
We restrict our attention to rods with isotropic bending behavior, $K_1=K_2=K_0$, where $\mathbf{U} = \mathbf{M}/K_0 + (1-K_3/K_0)\, U_3 \,  \mathbf{d_3}$ with $U_3 = \left( M_x \, d_{3x} + M_y \, d_{3y} + M_z \, d_{3z} \right)/K_3$.
In this case (\ref{equa_M3}) shows that $\d U_3 / \d S \equiv 0$ $\forall S$, and consequently $\mathbf{U}$ vanishes for the equations~\cite{Kehrbaum1997,neukirch+henderson:2002}.
We additionally focus here on the weightless, $\mathbf{W}=\mathbf{0}$, and symmetrical $S_A=L/2-\Sigma$,  $S_B=L/2 +\Sigma$ case.
We use the diameter $D=2R$ of the spherical drop as unit length, and the buckling load $K_0/D^2$ as unit force, that is we  introduce the following dimensionless quantities
\begin{subequations}
\label{eq:adim}
\begin{align}
s = \frac{S-L/2}{D} \; ; \quad 
\ell = \frac{L}{D} \; ; \quad
(x,y,z) = \frac{(X,Y,Z-(L-\Delta)/2)}{D}  \; ;  \\
\sigma = \frac{\Sigma}{D} \; ; \quad
k_3 = \frac{K_3}{K_0} \; ; \quad
\delta = \frac{\Delta}{D} \; ; \quad
\mathbf{u} = \mathbf{U} \, D \; ; \quad 
f_\gamma = \frac{F_\gamma \, D^2}{K_0} \; ;   \\
\mathbf{n} = \frac{\mathbf{N} D^2}{K_0}  \; ; \quad 
\mathbf{m} = \frac{\mathbf{M} \, D}{K_0}   \; ; \quad
(v,v_0) = \frac{(V_\mathrm{wall},V_0) \,D^2}{K_0} \, . 
\end{align}
\end{subequations}
We further restrict the study to equilibrium shapes being invariant by a rotation of angle $\pi$ about the line passing through the middle point of the rod, $\mathbf{r}(s=0)$, and directed along the axis $\mathbf{e_y}$. Such configurations are sometimes referred to as flip-symmetric \cite{domokos+healey:2001}. 
%, and in some cases it can be shown that only flip-symmetric solutions exist~\cite{neukirch+henderson:2002}.
%
The quantities $x(s)$, $z(s)$, $d_{1y}(s)$, $d_{2x}(s)$, $d_{2z}(s)$, $d_{3y}(s)$, $m_y(s)$, and $n_y(s)$ are then odd functions of $s$, while $y(s)$, $d_{1x}(s)$, $d_{1z}(s)$, $d_{2y}(s)$, $d_{3x}(s)$, $d_{3z}(s)$, $m_x(s)$, $m_z(s)$, $n_x(s)$, and $n_z(s)$ are even functions of $s$.
The center of the spherical drop consequently lies on the flip-symmetry axis, that is $x_C=0$ and $z_C=0$, and the end-rotation angles are such that $\alpha_0=-\alpha_L$.
Making use of this symmetry, we only integrate the equilibrium equations for $s \in [0, \ell/2]$, which read 
\begin{subequations}
\label{eq:18d-equilibre}
\begin{align}
x'(s)         = & \, d_{3x} \, , \; y'(s)         =  d_{3y} \, , \;  z'(s)         =  d_{3z} \, , \\
d_{1x}'(s) = & \, d_{1z} \, m_y - d_{1y} \, m_z + (1 - k_3) \, u_3 \, (d_{1z} \, d_{3y} - d_{1y} \, d_{3z}) \, , \\
d_{1y}'(s) = & \, d_{1x} \, m_z - d_{1z} \, m_x + (1 - k_3) \, u_3 \, (d_{1x} \, d_{3z} - d_{1z} \, d_{3x}) \, ,  \\
d_{1z}'(s) = & \, d_{1y} \, m_x - d_{1x} \, m_y + (1 - k_3) \, u_3 \, (d_{1y} \, d_{3x} - d_{1x} \, d_{3y}) \, , \\
d_{3x}'(s) = & \, d_{3z} \, m_y - d_{3y} \, m_z \, , \;  d_{3y}'(s) =  d_{3x} \, m_z - d_{3z} \, m_x \, , \;
d_{3z}'(s) = d_{3y} \, m_x - d_{3x} \, m_y \, , \\
m_x'(s)       = & \,  d_{3z} \, n_y - d_{3y} \, n_z \, , \;
m'_y(s)       =   d_{3x} \, n_z - d_{3z} \, n_x \, , \;
m'_z(s)       =   d_{3y} \, n_x - d_{3x} \, n_y  \, ,  \\
n_x'(s)     = & \, \chi \; \partial v/ \partial x  + 2 f_B \, x(\sigma)  \; \dirac_{s-\sigma}    \, ,   \\
n_y'(s)     = & \, \chi \; \partial v/ \partial y  + 2 f_B \, \left[ y(\sigma)-y_C \right]  \; \dirac_{s-\sigma}   \, ,  \\
n_z'(s)     = & \, \chi \; \partial v/ \partial z  + 2 f_B \, z(\sigma)  \; \dirac_{s-\sigma}  \, ,
%n_x'(s)     = & \, \chi \frac{\partial v}{\partial x} + 2 f_A \, x(s_A)  \, \delta(s-s_A) + 2 f_B \, x(s_B)  \, \delta(s-s_B)        \\
%n_y'(s)     = & \, \chi \frac{\partial v}{\partial y} + 2 f_A \, \left[ y(s_A)-y_C \right]  \, \delta(s-s_A)   + 2 f_B \, \left[ y(s_B)-y_C \right]  \, \delta(s-s_B)     \\
%n_z'(s)     = & \, \chi \frac{\partial v}{\partial z} + 2 f_A \, z(s_A)  \, \delta(s-s_A) + 2 f_B \, z(s_B)  \, \delta(s-s_B) 
\end{align}
\end{subequations}
where $\dirac_{s-s_0}$ is the Dirac distribution centered on $s_0$, and $v = v_0 \, \left( 1 + \rho - 2 \sqrt{x^2 + (y-y_C)^2 + z^2} \right)^{-1}$. For $s \in [0 ; \sigma)$ the rod lies inside the spherical drop and we have $\chi=1$, otherwise $\chi=0$.
We set $\alpha=0$ in~(\ref{sys:BC}) and consider $v_0$, $\rho$, $f_\gamma$, $\ell$, and $k_3$ as fixed parameters.
We look for equilibrium solutions by integrating (\ref{eq:18d-equilibre}) with initial conditions
\begin{subequations}
\label{eq:initcond}
\begin{align}
x(0)=& \, 0 \; , \; 
y(0)=y_0 \, , \; 
z(0)=0 \, , \\
d_{3x}(0)=& \,  \sin \theta_0 \, , \; 
d_{3y}(0)=0 \, , \; 
d_{3z}(0)=\cos \theta_0 \, , \\
d_{1x}(0)=& \,  \cos \theta_0  \, , \;
d_{1y}(0)=0 \, , \; 
d_{1z}(0)=-\sin \theta_0 \, , \\
n_{x}(0)=& \,  n_{x0}  \, , \;
n_{y}(0)=0 \, , \; 
n_{z}(0)= n_{z0} \, , \\
m_{x}(0)=& \,  m_{x0}  \, , \;
m_{y}(0)=0 \, , \; 
m_{z}(0)= m_{z0}  
\end{align}
\end{subequations}
where $y_0$, $\theta_0$, $n_{x0}$, $n_{z0}$, $m_{x0}$, $m_{z0}$ along with $y_C$, $f_B$, and $\sigma$ are 9 unknowns which are balanced by the following 8 conditions.
We restrict to cases where the director $\mathbf{d_1}$ is aligned with the $\mathbf{e_x}$ axis at both extremities of the rod, that is we set  $\alpha_0=\alpha_L=0$.
Then, using (\ref{sys:BC}), we write 5 boundary conditions for the right extremity of the rod, $s=\ell/2$
\begin{equation}
x(\ell/2)=0 \; ; \; y(\ell/2)=0 \; ; \; d_{3x}(\ell/2)=0 \; ; \; d_{3y}(\ell/2)=0 \; ; \; d_{1y}(\ell/2)=0  \; ; 
\end{equation}
At the meniscus point $B$, we have 3 conditions, adapted from~(\ref{equa:AetBsphere}), (\ref{equa:saut_global}), and (\ref{equa:fb})
\begin{subequations}
\label{eq:bcnd-end}
\begin{align}
\sqrt{x^2(\sigma) + (y(\sigma)-y_C)^2 + z^2(\sigma)} = & \, 1/2 \, , \\
n_y(\sigma)= & \, 0 \, , \\
2 f_B \left[ -x(\sigma) \, d_{3x}(\sigma) +(y_C-y(\sigma))\, d_{3y}(\sigma) -z(\sigma) \, d_{3z}(\sigma) \right] = &\,  -f_\gamma+ v(\sigma) \, .
\end{align}
\end{subequations}
The solution set is thus a $9-8 = 1$ dimensional manifold and we plot in Section \ref{section:bif-diag} different solution paths.

\section{Bifurcation diagram} \label{section:bif-diag}
%===================================
%
%
%
%
%
%
%
%
%
%
%
%
\begin{figure}[ht]
\centering
\includegraphics[width=.98\columnwidth]{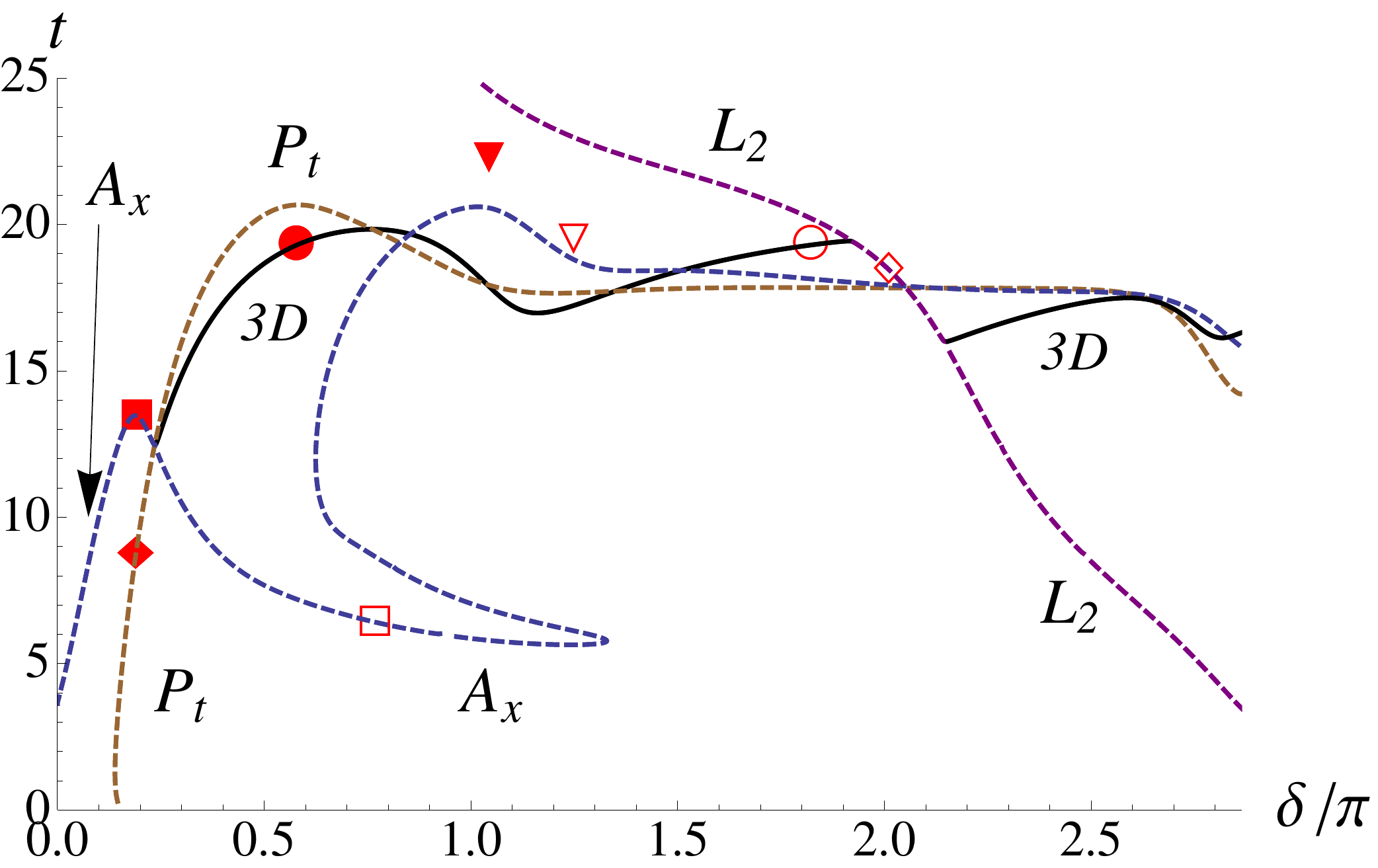}
\caption{Force-displacement post-buckling curves for $f_\gamma=20$ and ${\cal R}=0$, where $t$ is the applied tension and $\delta$ the end-shortening. Paths of 3D and 2D configurations are plotted with solid and dashed lines respectively. Points marked with empty and filled circles, squares, diamonds, and triangles correspond to configurations shown in Figures~\ref{fig:povray1}, \ref{fig:povray2},  \ref{fig:povray3}, and \ref{fig:povray4} respectively. Points marked triangles belong to paths not shown here.}
\label{fig:bif-path-fg-20}
\end{figure}

We numerically solve the boundary-value problem defined in Section \ref{section:bvp}, using either a shooting method or the AUTO collocation method \cite{doedel:1991}. Pseudo-arc-length continuation then enables us to follow the solutions as parameters are varied. In order to compare present three-dimensional (3D) results with two dimensional (2D) solutions studied in \cite{Elettro2015}, we use the same parameters $k_3=0.9$, $\ell=10$, $f_\gamma=20$, and $v_0 = 0.02 / \ell^2=2 \times 10^{-4}$, but  reduce $\rho$ to $\rho=0.1$ thereby bringing the meniscus points closer to the bounding sphere.
% In addition to the  planar equilibrium paths computed in \cite{Elettro2015} we here report on a path comprising 3D configurations.

We start with a straight configuration subject to a large applied tension $t = \mathbf{n}(\ell/2) \cdot \mathbf{d_3}(\ell/2)$.
\begin{figure}[ht]
\centering
\includegraphics[width=.98\columnwidth]{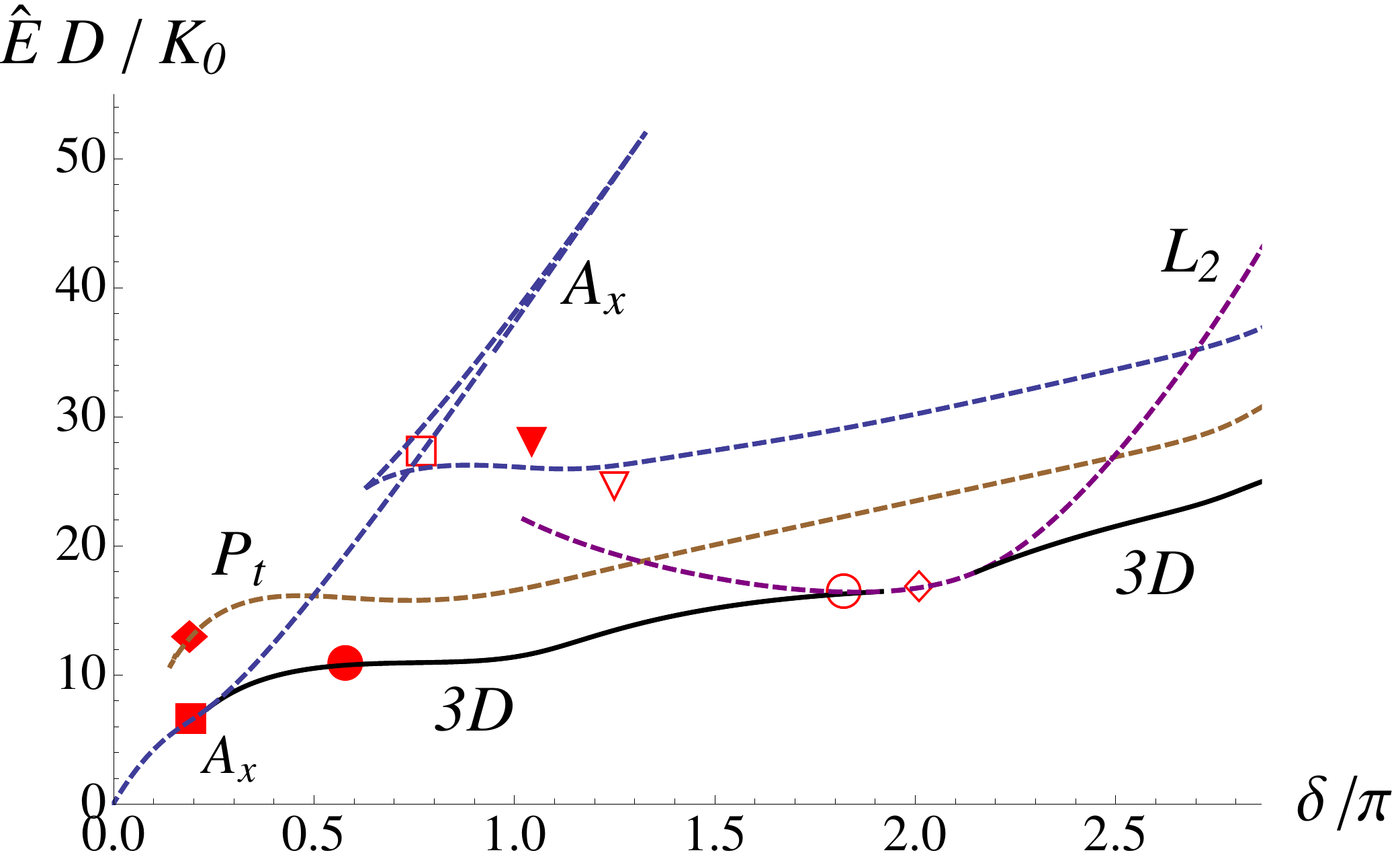}
\caption{Energy as function of the end-shortening $\delta$ for the post-buckling curves of Figure \ref{fig:bif-path-fg-20}.}
\label{fig:bif-path-fg-20E}
\end{figure}
We define the end rotation ${\cal R}$ as the angle between $\mathbf{d_1}(\ell/2)$ and $\mathbf{d_1}(-\ell/2)$, 
${\cal R} = \alpha_L-\alpha_0 = 2 \, \alpha_L$
%that is twice the angle between $\mathbf{d_1}(\ell/2)$ and  $\mathbf{e_x}$, 
and we first restrict to configurations with ${\cal R}=0$.  The straight configuration is consequently twistless. (For the present case of clamped-clamped configurations this end-rotation ${\cal R}$ is closely related to the topological link $Lk$, see {\em e.g.} \cite{Peletier2007}.) As the tension $t$ is decreased under a threshold value $t \simeq 3.7$, the rod buckles and the post-buckling regime first involves 2D configurations, this is the path $Ax$ introduced in  \cite{Elettro2015}.
\begin{figure}[ht]
\centering
\includegraphics[width=.47\columnwidth]{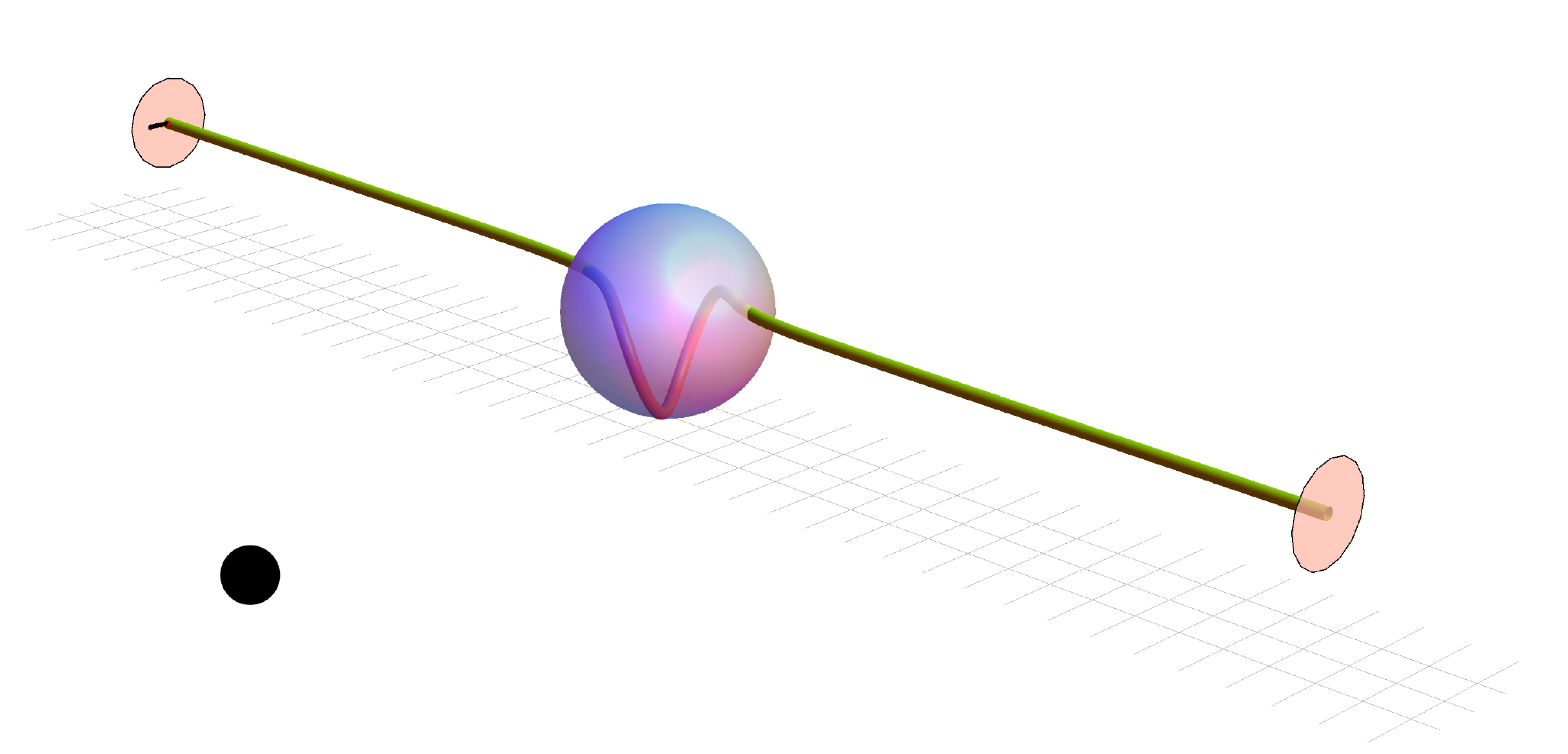}
\hspace{0.01cm}
\includegraphics[width=.48\columnwidth]{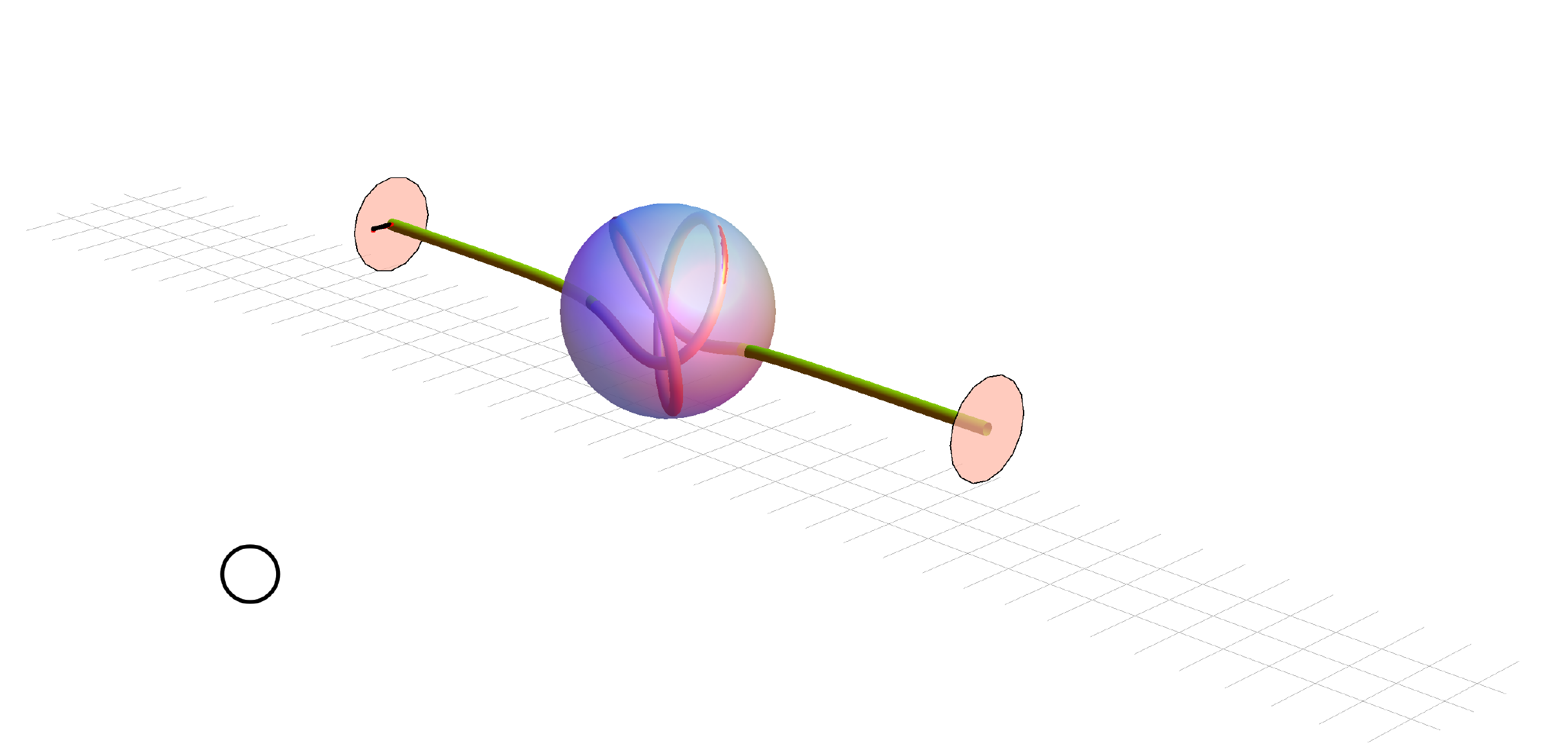}
\caption{3D configurations corresponding to the filled and empty circles on the post-buckling diagram of Figures~\ref{fig:bif-path-fg-20} and \ref{fig:bif-path-fg-20E}.}
\label{fig:povray1}
\end{figure}
The dimensionless end-shortening $\delta$ is then gradually increased, and at $\delta \simeq 0.24 \,  \pi$ lies a pitchfork bifurcation and a secondary path, consisting of 3D configurations, emerges and progresses toward the plateau value $t_\text{P}=f_\gamma-2$~\cite{Elettro2015a}, see Figure~\ref{fig:bif-path-fg-20}. Later along the path, for $\delta \simeq 1.9 \, \pi$, another pitchfork bifurcation occurs and, for a limited $\delta$ interval, the equilibrium configurations become planar again: The 3D path merges with the path $L_2$, which consists in configurations which are planar and looping twice inside the sphere.
\begin{figure}[ht]
\centering
\includegraphics[width=.47\columnwidth]{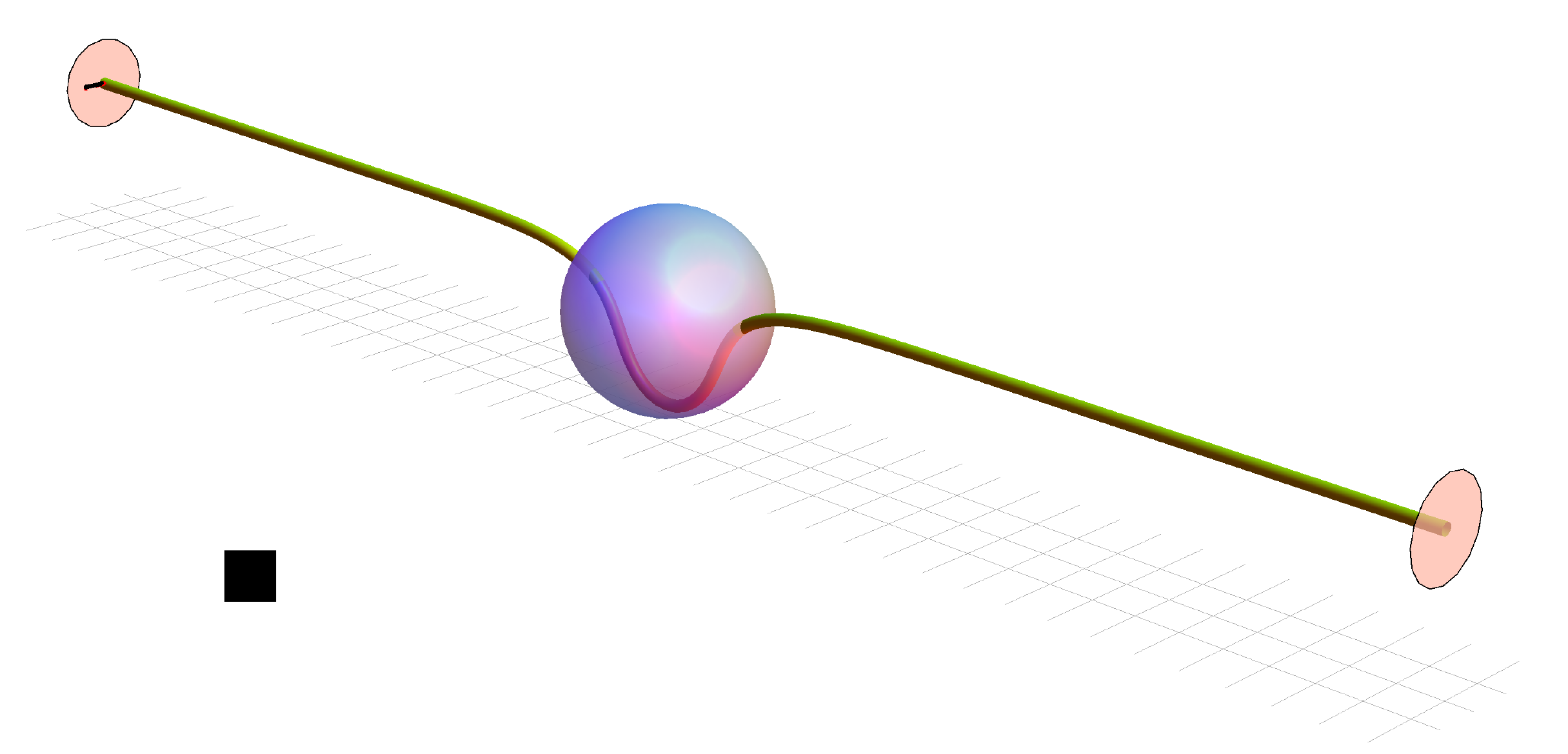}
\hspace{0.01cm}
\includegraphics[width=.48\columnwidth]{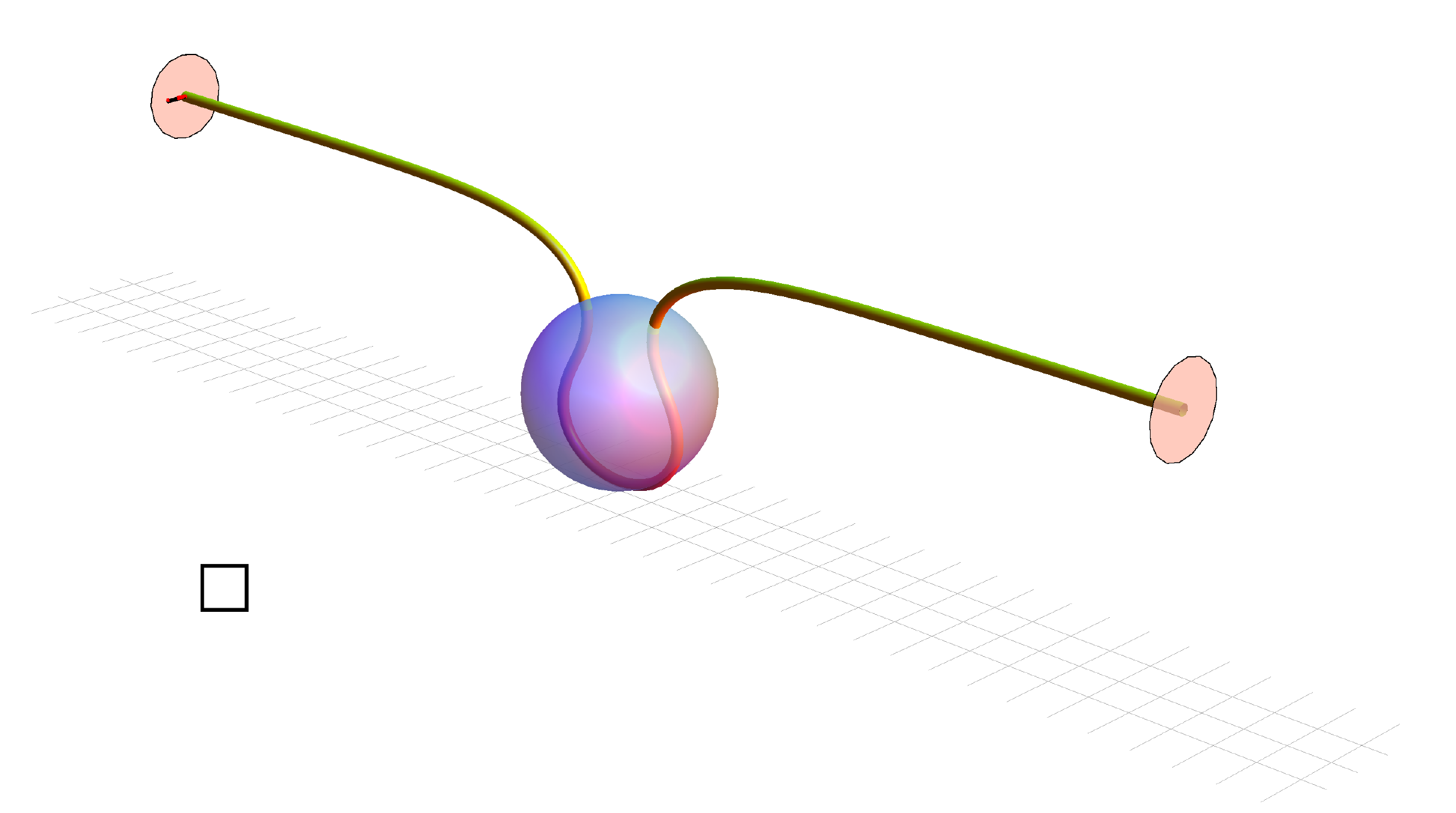}
\caption{2D configurations corresponding to the filled and empty squares on the post-buckling diagram of Figures~\ref{fig:bif-path-fg-20} and \ref{fig:bif-path-fg-20E}.}
\label{fig:povray2}
\end{figure}
After yet another pitchfork bifurcation at $\delta \simeq 2.15 \,  \pi$, the $L_2$ and 3D paths split again. 

\begin{figure}[ht]
\centering
%=============================================
\includegraphics[width=.95\columnwidth]{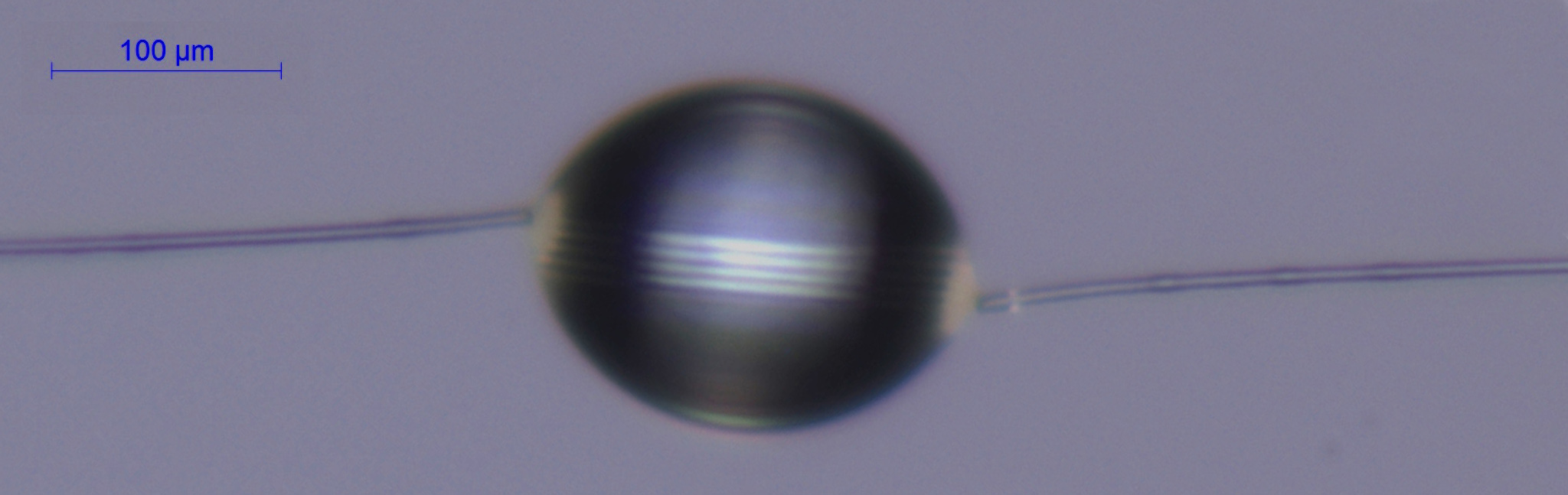}\\
% On voit la fibre passer 4 fois donc il y a 4.5 tours en fait, cad 2.83 mm de fibre en goutte.
% Comme d’habitude, c’est du TPU avec de l’huile silicone, on a donc :
% E  = 17MPa
% \gamma = 21.1 mN/m
% \theta = 23 deg
% h = 5.6 +/- 0.5 microns (diamètre fibre)
% D = 200.2 microns (diamètre goutte)
%
%=============================================
\vspace{0.5cm}
%=============================================
\includegraphics[width=.95\columnwidth]{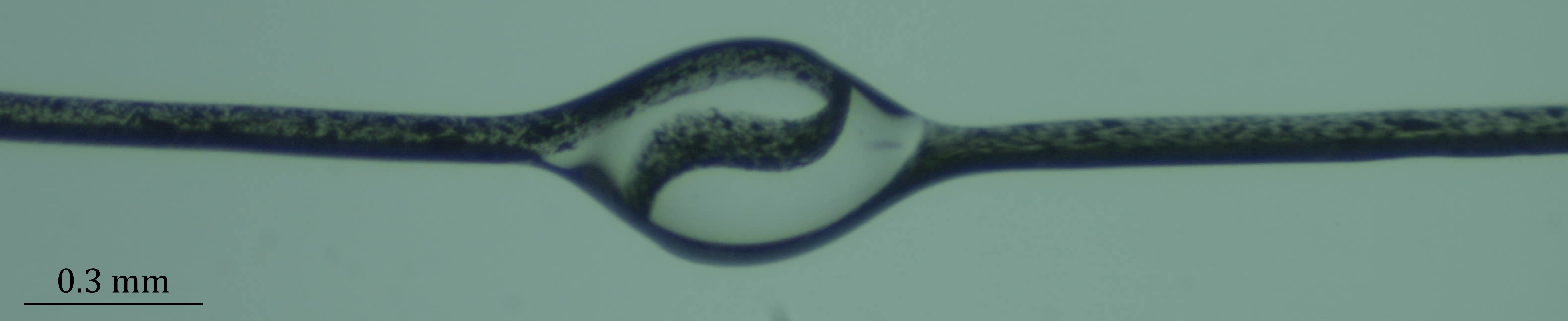}
%fibre en forme de calotte cylindrique (~environ une lamelle) avec w=160um, h=40um en PVS8 dureté E~200kPa. 
%Goutte d'huile silicone gamma~0.021N/m (rho=960kg/m3). 
%Longueur mouillée initiale = 0.71mm (car la goutte n'est pas sphérique, elle est allongée). 
%=============================================
\caption{(Up) Thermoplastic Poly Urethane (Young's modulus 17 MPa) fiber with circular cross-section (diameter 5.6 +/- 0.5 microns) spooled in a silicone oil drop (diameter $\simeq$ 200 microns, surface tension 21 mN/m) with ${\cal R}=0$. There is $2 \Sigma = 2.8$ mm of fiber in the drop, which corresponds to $\simeq$ 4.5 spools. (Down) PolyVinylSiloxane beam (Young's modulus $\simeq$ 200kPa) with rectangular cross-section (40 $\times$ 160 microns) bent in a silicone oil drop (diameter $\simeq$ 0.7 mm) with ${\cal R}=0$. The 2D configuration obtained, analogous to configurations along the $P_t$ curve in Figure~\ref{fig:bif-path-fg-20}, is due to the anisotropy of the cross-section which strongly favors bending in one direction.
}  
\label{fig:photo}
\end{figure}

Another path, called $Pt$ in \cite{Elettro2015}, exists and is composed of configurations which are always planar. These configurations, resembling a second buckling mode, are probably unstable, but using a rod with a flat, rectangular cross-section we were able to stabilize them, see Figure~\ref{fig:photo}. No bifurcation was found along this path $P_t$.
\begin{figure}[ht]
\centering
\includegraphics[width=.47\columnwidth]{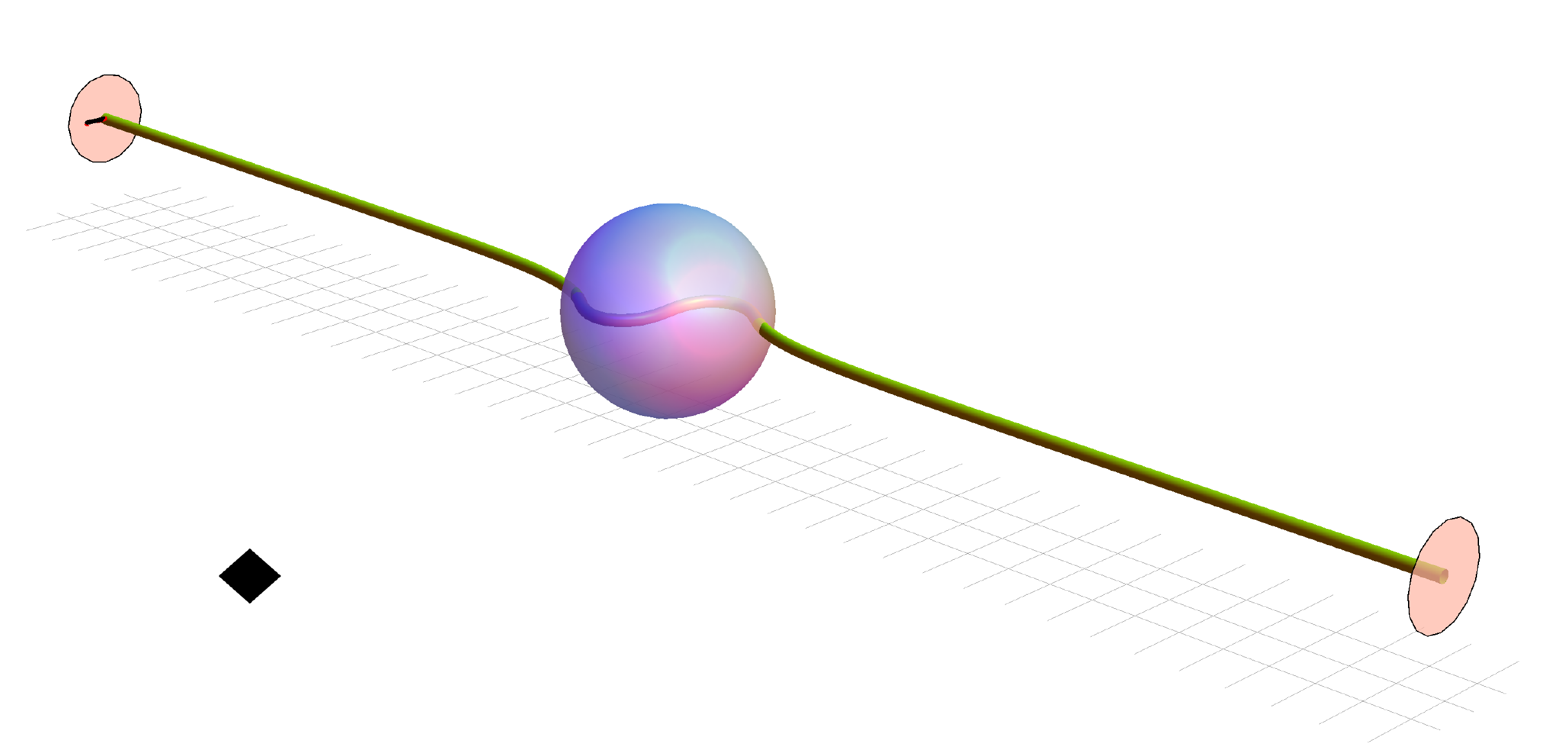}
\hspace{0.01cm}
\includegraphics[width=.48\columnwidth]{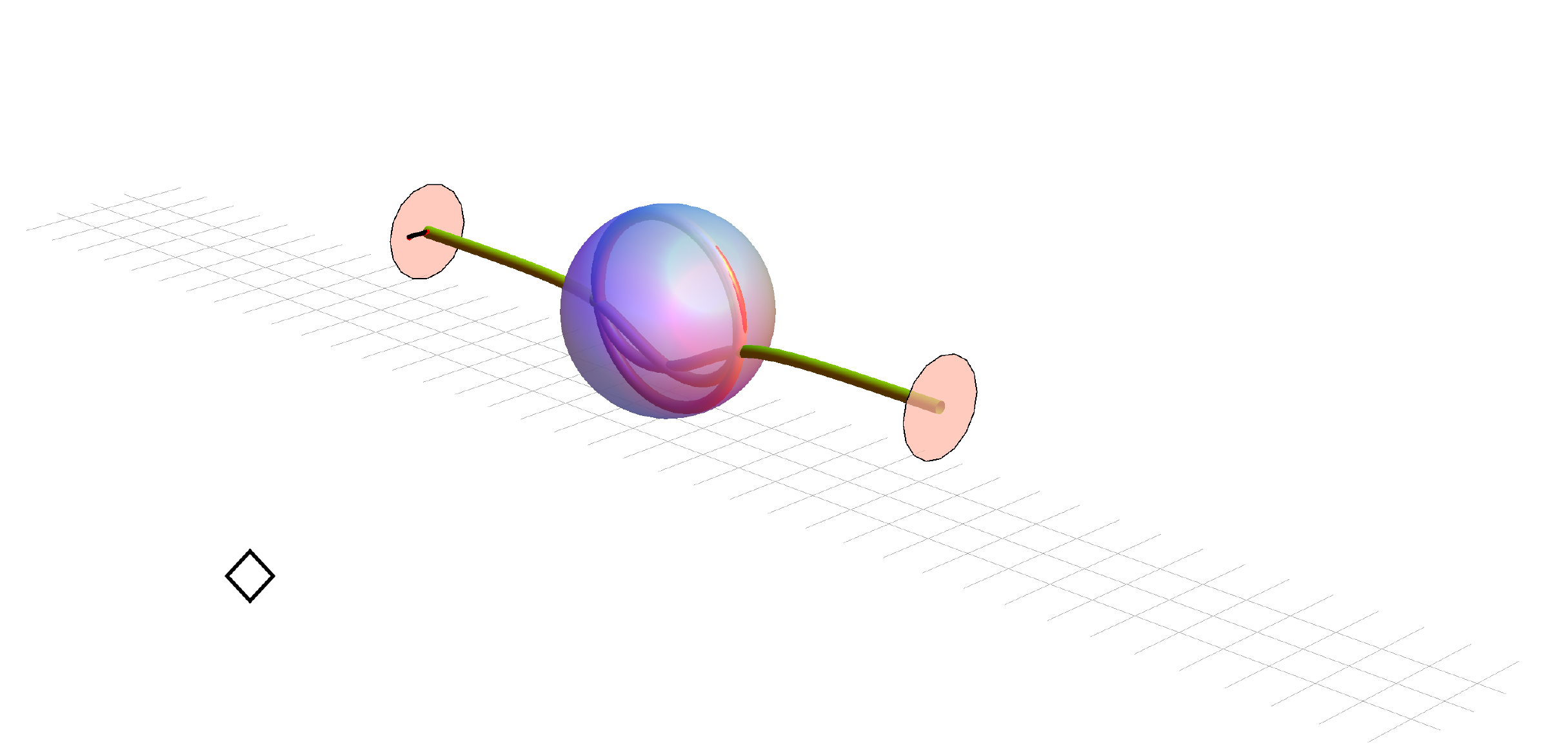}
\caption{2D configurations corresponding to the filled and empty diamonds on the post-buckling diagram of Figures~\ref{fig:bif-path-fg-20} and \ref{fig:bif-path-fg-20E}.}
\label{fig:povray3}
\end{figure}

\begin{figure}[ht]
\centering
\includegraphics[width=.47\columnwidth]{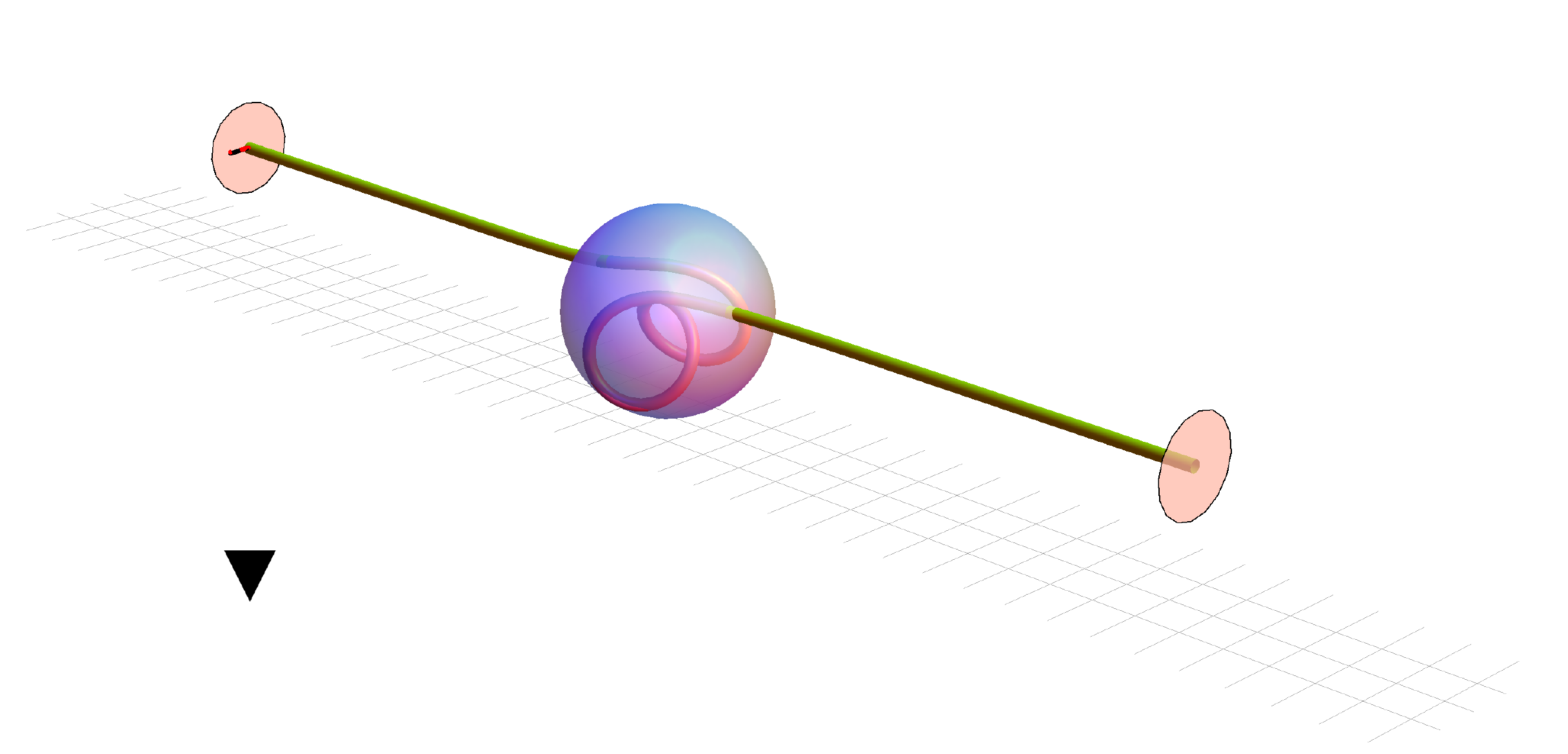}
\hspace{0.01cm}
\includegraphics[width=.48\columnwidth]{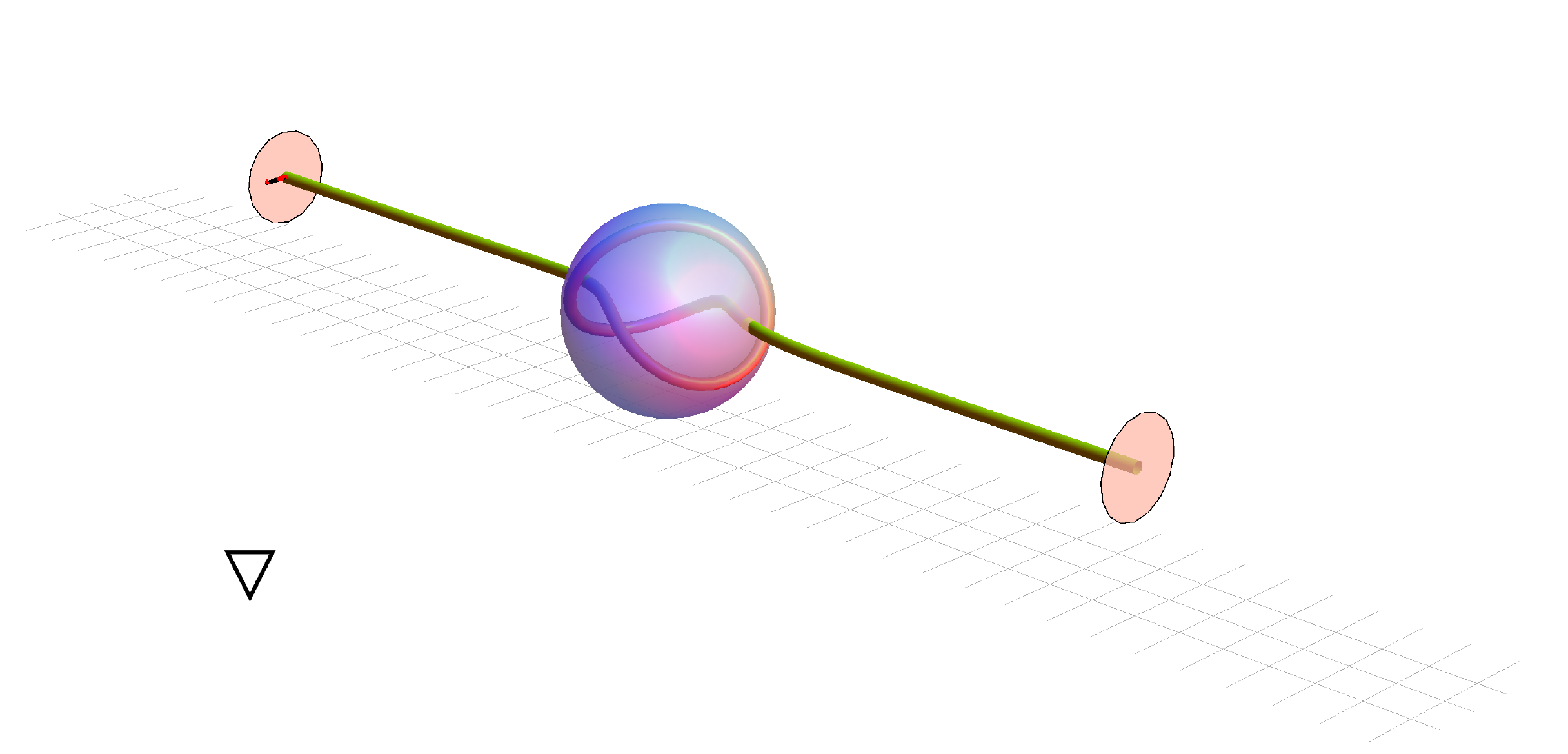}
\caption{3D configurations corresponding to the up and down triangles on the post-buckling diagram of Figures~\ref{fig:bif-path-fg-20} and \ref{fig:bif-path-fg-20E}.}
\label{fig:povray4}
\end{figure}

\begin{figure}[ht]
\centering
\includegraphics[width=.67\columnwidth]{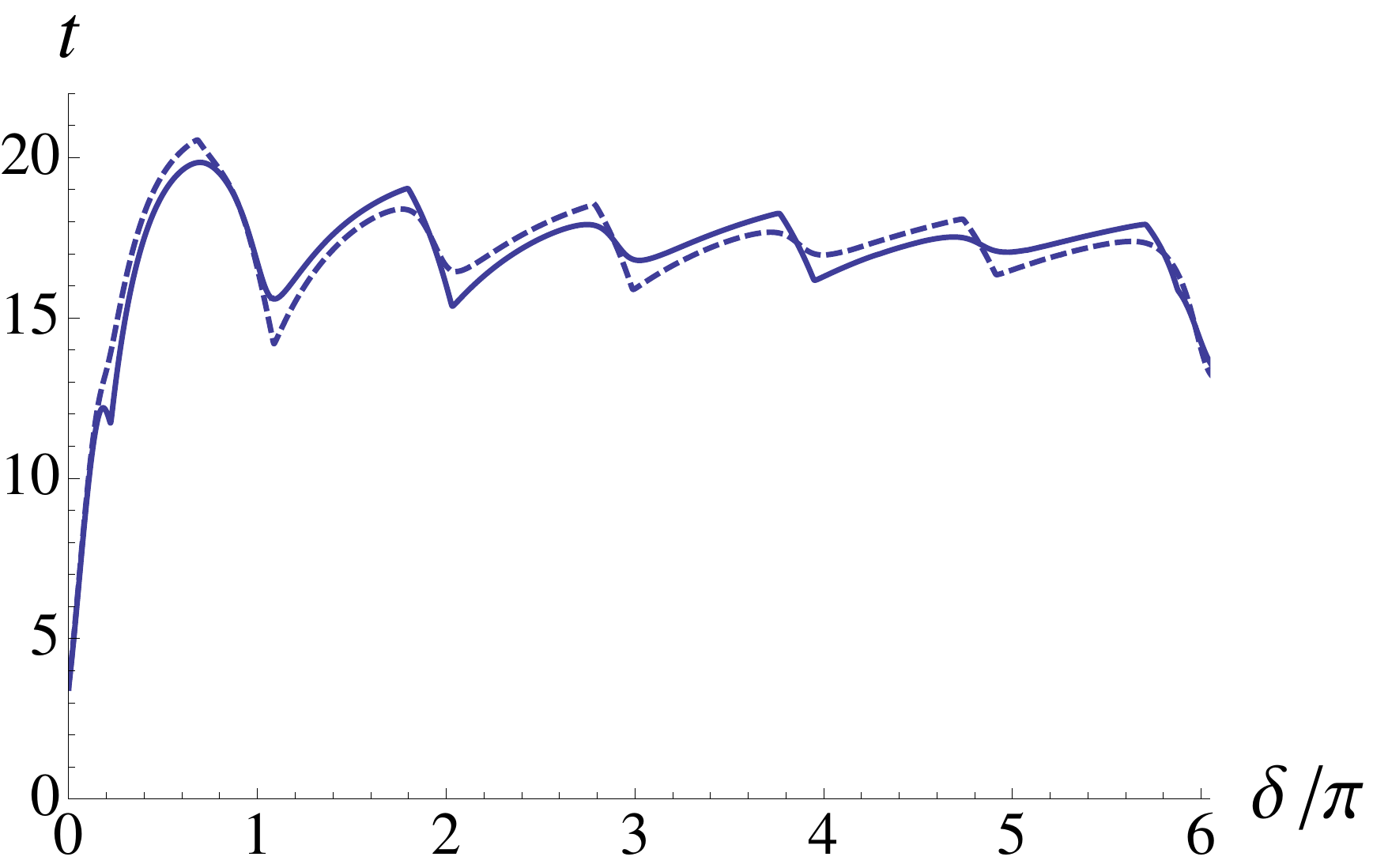}
\caption{Force-displacement post-buckling curves for $\ell=20$, $f_\gamma=20$, and either ${\cal R}=0$ (continuous line) or ${\cal R}=2 \pi$ (dashed line). Only the succession $Ax$-3D-$L_2$-3D, which has lowest energy, is shown. For ${\cal R}=0$ (${\cal R}=2 \pi$), the series of events of planar-loop configurations appears around even (odd) values of $\delta/\pi$.}
\label{fig:Lk0_et_Lk1}
\end{figure}

In order to classify these different paths we plot the energy $\hat{E}=E_\text{tot}+ F_\gamma (\Delta+D)$ in Figure~\ref{fig:bif-path-fg-20E} where the lowest energy path is seen to be the succession $Ax$-3D-$L_2$-3D. Please note that yet other paths were found, but with higher energies, see {\em e.g.} the two configurations in Figure~\ref{fig:povray4}.
In \cite{Elettro2015} a path $L_1$, comprising 2D configurations which were looping once inside the sphere, was shown. This path does not fulfill the topological constraint ${\cal R}=0$ and is therefore not plotted in Figures~\ref{fig:bif-path-fg-20} or \ref{fig:bif-path-fg-20E}. Nevertheless if the boundary condition ${\cal R}=2\pi$ was used instead, such a path $L_1$ would  come into play. We draw in Figure~\ref{fig:Lk0_et_Lk1} the post-buckling of a rod with $\ell=20$ (and $k_3=0.9$, $f_\gamma=20$, $v_0 = 2 \times 10^{-4}$,  $\rho=0.1$) and either ${\cal R}=0$ or ${\cal R}=2 \pi$.
We see that for ${\cal R}=0$ (${\cal R}=2 \pi$), around $\delta / \pi = 2 j$ ($2 j-1$) there is an interval in which configurations with $2j$ ($2j-1$) loops exists, with $j=1,2, 3, \ldots$. As the system tries to minimize the bending energy, the rod tends to wind in the sphere following the largest possible radius of curvature. This $\delta = \pi$, or $\Delta = 2\pi \, R$, periodicity consequently corresponds to the addition of one coil (of radius $R$) in the sphere: as the rod enters the attracting sphere there are periodic events where the rod adopts an ordered configuration resembling a spool. We have been able to experimentally evidence these ordered states, see Figure~\ref{fig:photo} and \cite{elettroW4}.

\section{Conclusion}
%==============
%
%
%
%
%
%
%
%
%
%
%
In conclusion we have presented a model for the interaction of an elastic rod with a liquid drop, with the restriction that the drop remains spherical. The difference in surface energies $\gsv-\gsl$ yields a capillary force that compresses the part of the beam which lies inside the drop. When the compression is large enough the rod buckles and coils in the drop. We have derived the rod's equilibrium equations from a variational point of view, showing that the compressive forces applied on the rod at the meniscus points were oriented toward the center of the spherical drop (Eq.~\ref{sys:meniscus-force}) and had their intensity depending of their orientation relative to the rod's tangent (Eq.~\ref{sys:resolveFAB}).
We have numerically solved the equilibrium equations and found planar and spatial coiled configurations, with bifurcations between them as the end-shortening of the system is increased. More precisely we found that there is an interplay between 2D and 3D solutions, the lowest energy solution being mainly 3D with short intervals in which the rod adopts a planar-loop configuration. This intermittency scenario has still to be verified experimentally \cite{elettroW4} but we show in Fig.~\ref{fig:photo} that solutions where the rod is tidily spooled inside the drop indeed exist.

\begin{acknowledgements}
%===================
%
%
%
%
We thank Camille Dianoux and Sinan Haliyo for their help on microscopy, and Arnaud Antkowiak for comments on the variational approach.
The present work was supported by ANR grant ANR-09-JCJC-0022-01, ANR-14-CE07-0023-01, and ANR-13-JS09-0009. Financial support from `La Ville de Paris - Programme \'Emergence' and CNRS, through a PEPS-PTI grant, is also gratefully acknowledged.
\end{acknowledgements}

%\section*{Appendix: Derivation of the equilibrium equations}
%=============================================

\bibliographystyle{spmpsci}      % mathematics and physical sciences
\bibliography{windlass3}   % name your BibTeX data base

\end{document}